\documentclass{PoS}
\title{Three-body forces in Bethe-Salpeter and light-front equations}

\ShortTitle{Three-body forces in Bethe-Salpeter and light-front
equations}

\author{\speaker{V.A.~Karmanov}
         \\
        Lebedev Physical Institute, Moscow, Russia\\
        E-mail: \email{karmanov@sci.lebedev.ru}}

\author{P.~Maris\\
Iowa State University, Ames, Iowa, USA}

\abstract{In relativistic frameworks, given by the Bethe--Salpeter
and light-front bound state equations, the binding energies of
system of three scalar particles interacting by scalar exchange
particles are calculated. In contrast to two-body systems, the
three-body binding energies obtained in these two approaches
differ significantly from each other: the ladder kernel in
light-front dynamics underbinds by approximately a factor of two
compared to the ladder Bethe--Salpeter equation. By taking into
account three-body forces in the light-front approach, generated
by two exchange particles in flight, we find that most of this
difference disappears; for small exchange masses, the obtained
binding energies coincide with each other.}

\FullConference{LIGHT CONE 2008 Relativistic Nuclear and Particle Physics\\
                 July 7-11, 2008\\
                 Mulhouse, France}

\begin{document}

\section{Introduction}
In this article we sketch out the results of our theoretical study
\cite{km08} of three-body bound states, carried out in full
relativistic framework, given by the Bethe--Salpeter (BS) approach
\cite{SB_51} and light-front dynamics (LFD) \cite{cdkm,bpp}.
Eventually, the goal is to investigate the bound states of
fermions, using e.g. a meson-exchange model in the case of
3-nucleon bound states (triton, ${}^3$He), and baryons as bound
states of three (non-perturbatively dressed) quarks interacting
via gluons, and calculate their electromagnetic form factors. For
simplicity however, and as a first step, we restrict ourselves to
spinless system. For such system, the three-body BS equation with
one-boson exchange interaction was first solved and the results
were presented in \cite{Pieter}. As far as we know, the LFD
three-body equation with one-boson exchange kernel has not yet
been solved. Previously, the three-body LFD equation was solved in
\cite{tobias,ck_03} for zero-range interaction.  Its solution was
also found in the relativistic quantum-mechanical approach for
bound and scattering states \cite{polyzou}, with a
phenomenological mass operator.

\begin{figure}[!ht]
\includegraphics[width=7cm]{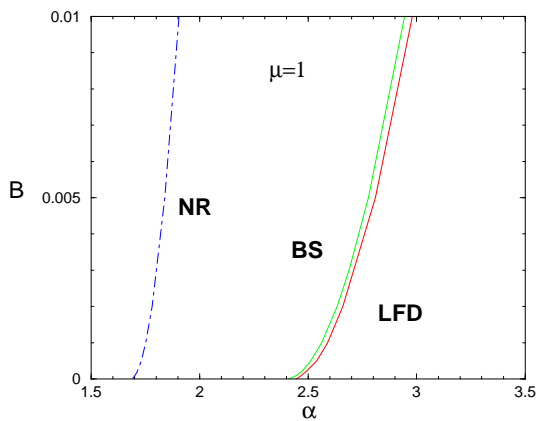}\caption{Binding
energy $B$ of system of two spinless particles interacting by
one-boson exchange with mass $\mu=1$ v.s. the coupling constant
$\alpha=g^2/(16\pi m^2)$, calculated by BS, LFD and Schr\"odinger
equations. The units are set by $m=1$. The figure is taken from
\cite{mc_2000}.} \label{mc}
\end{figure}

Though a system of spinless particles is oversimplified, some its
properties are rather general and their study gives useful
physical insight. Especially, the comparison of results found in
the BS and LFD approaches is fruitful. An important observation is
the fact that for two-body system, these two approaches give
results very close to each other both for the binding energy
\cite{mc_2000} and also for electromagnetic form factors
\cite{ckm08}. The comparison of BS and LFD calculations, for the
constituent mass $m=1$ and for the exchange mass also $\mu=1$, is
shown in fig. \ref{mc}. The curve NR means non-relativistic
binding energy (obtained via Schr\"odinger equation), which
considerably differs, even for weakly bound system, from almost
coinciding with each other BS and LFS binding energies.

It is important to note, however, that the actual ladder kernels
in the BS approach and in LFD are not identical, nor are they
given by the same graphs.  In the BS ladder kernel, there is no
notion of time-ordering in the diagrams.  On the other hand, the
kernel for the LFD equation is given by the time-ordered graphs in
the light-front (LF) time.
\begin{figure*}[htbp]
\begin{center}
\begin{minipage}{14.cm}
\mbox{\includegraphics[width=5cm]{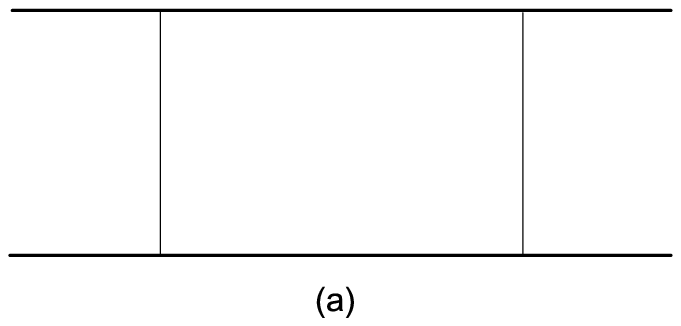} \hspace{0.5cm}
\includegraphics[width=5cm]{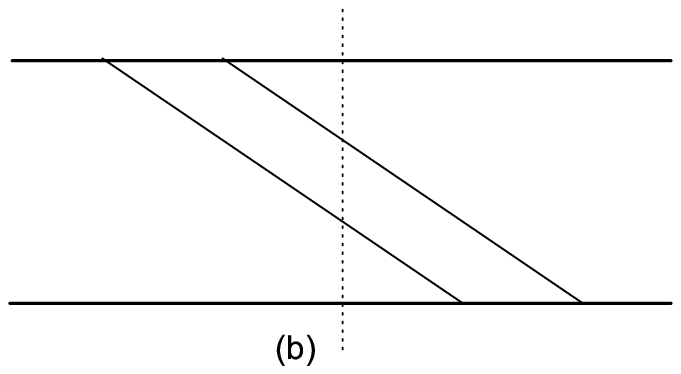}}
\end{minipage}
\end{center}
\caption{(a) Feynman ladder graph with two exchanges.  (b) One of
six time-ordered graphs ("stretched box"), generated by the ladder
Feynman graph (a). \label{feyn2}}
\end{figure*}
Thus, the second iteration in the two-body ladder BS equation (the
second-order Feynman box graph, figure~\ref{feyn2}(a)), when
represented as a set of time-ordered graphs, turns into six LF
time-ordered graphs, including two so-called ``stretched boxes''
with two exchange particles in the intermediate state.  One of
such stretched boxes is shown in figure~\ref{feyn2}(b).  These
stretched boxes with two (and more) exchange particles in the
intermediate state are implicitly included in the BS equation, but
they are not generated by iterations of the LF ladder kernel;
therefore, they are omitted in the corresponding LFD bound state
equation.  In principle, this will cause a difference between the
BS and LFD results.  However, direct examination of the stretched
boxes \cite{sbk,bs2} shows that they are small. This explains, why
the BS and LFD two-body binding energies are very close to each
other, despite the fact that the kernels are not identical.

We will see below that in the three-body problem the situation is
quite different: even with a simple one-boson exchange kernel, the
binding energies obtained by BS and LFD significantly differ from
each other, in drastic contrast to two-body case. This difference
will be explained by the three-body forces generated by two
exchange particles in fight.

\section{Three-body results for ladder kernel}
\begin{figure}[!ht]
\center
\includegraphics [width=7cm]{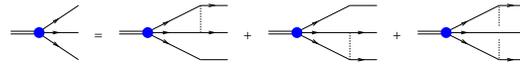}
\caption{Three-body bound state equation with ladder kernel.
\label{3bodyeq}}
\end{figure}

We consider a scalar field theory, with interaction $-g\phi^2
\varphi$, and the bound states consisting of three particles
$\phi$, interacting via the exchange of a $\varphi$ boson.
Three-body equation is graphically shown in fig. \ref{3bodyeq}.

The one-boson exchange BS kernel is simply
$$
K(k,k')=\frac{-g^2}{(k-k')^2-\mu^2+i\epsilon}\ .
$$
In LFD, this kernel is represented through LF variables
$\vec{k}_{\perp},x$ (see e.g. eq. (7.27) from \cite{cdkm}).
Exactly the same kernels were used in the two-body BS and LFD
equations. On energy shell, the expressions for BS and LFD
kernels, being represented in terms of the same variables, are
identical. Whereas off energy shell they differ from each other.

The three-body BS amplitude $\Gamma(p,q;P)$ ($p$ and $q$ are the
Jacobi four-momenta) depends on five scalar variables: $p^2,\;
q^2,\; p\cdot q,\; p\cdot P,\; q\cdot P$. The square $P^2=M^2$  is
fixed by the value of the bound mass $M$. To solve the equation
numerically, we transform it to Euclidean space.

To parametrize the three-body LF wave function
$\psi(\vec{k}_{12\perp},x_{12};\vec{k}_{3\perp},x_3)$, we
introduce transverse momenta $\vec{k}_{\perp}$ and the ratios $x$
and construct from them  LF Jacobi variables:
$\vec{k}_{12\perp},x_{12}$ are defined for two-body subsystem 12,
$\vec{k}_{3\perp},x_3$ are momenta of the 3rd particle.  The LF
wave function also depends of five scalar variables:
$\vec{k}^2_{12\perp},\; \vec{k}^2_{3\perp},\;
\mbox{$\vec{k}_{12\perp}\cdot\vec{k}_{3\perp}$},$ $x_{12},\; x_3$.

Thus we have an integral equation, which is four-dimensional in
the BS approach and  three-dimensional in LFD, for the bound state
amplitude which is a function of five independent variables in
both cases. We solve it numerically by straightforward
discretization of the internal and external variables; we do not
make use of any expansion of $\Gamma$ and $\psi$ in partial waves,
nor any other set of basis functions.

\begin{figure}[tb]
\includegraphics[width=.6\textwidth]{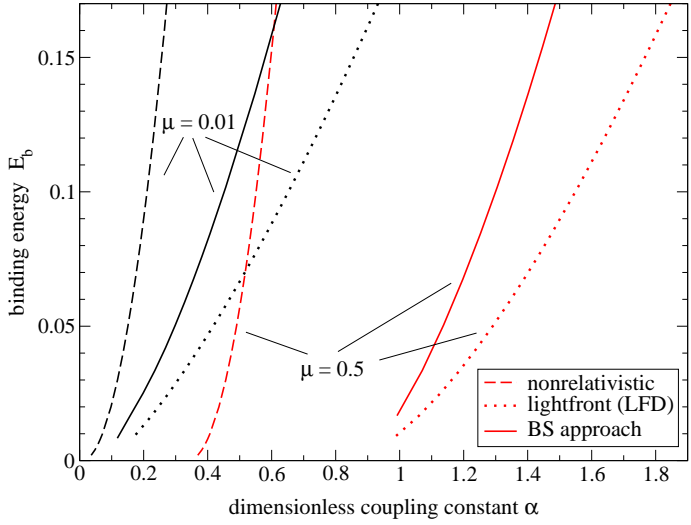}
\caption{Three-body binding energy for one boson exchange with
masses $\mu=0.01$ and $\mu=0.5$ (units set by $m = 1$).  Solid
curves are the BS result; dotted curves are the LFD results;
dashed curves are the solution of the non-relativistic
Schr\"odinger equation.\label{fig5}}
\end{figure}

For small binding energies $E_b=3m-M_3$, and for two exchange
masses $\mu=0.01$ and $\mu=0.5$, our results are shown in
figure~\ref{fig5} (the units are set by $m=1$).  For comparison,
we also include the non-relativistic result obtained by solving
the non-relativistic Faddeev equation in the momentum space.
Clearly, all three calculations give different results. We
emphasize the large difference between BS and LFD three-body
binding energies, in contrast to coincidence of the two-body ones,
shown in fig. \ref{mc}.

\begin{figure}[!ht]
\centering
\includegraphics[width=.6\textwidth]{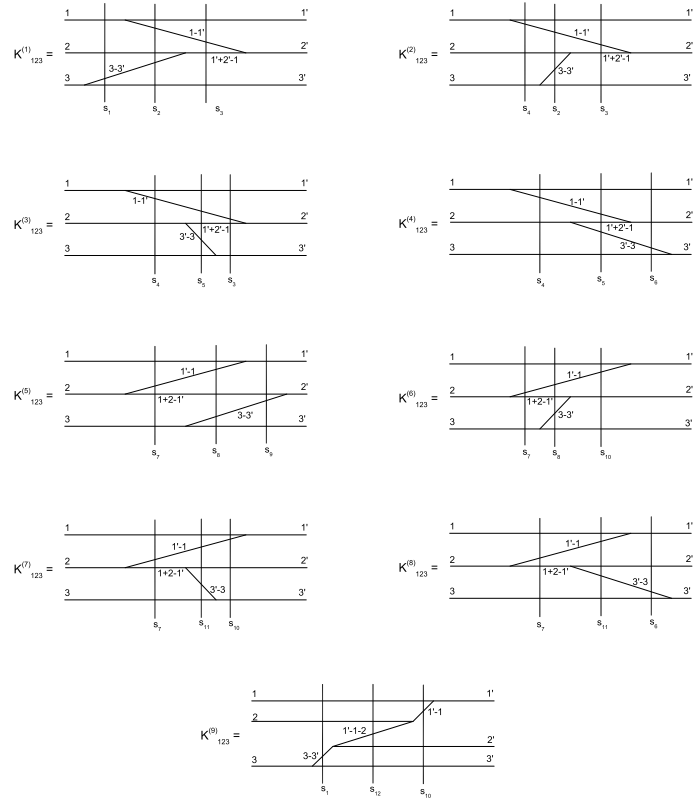}
\caption{Contribution to three-body forces from two bosons in
flight ($K^{(1-8)}_{123}$) and from pair-creation
($K^{(9)}_{123}$).} \label{3bodyf}
\end{figure}

In this way, new interesting and important features emerge,
highlighting differences between the BS approach and LFD. In a
three-body problem the difference between LFD and time-ordered
iterated BS kernels cannot be reduced to (negligible) stretched
boxes: new LFD (non-loop) diagrams appear, which are shown in
figure~\ref{3bodyf}. These graphs correspond to three-body forces
which are not taken into account in the three-body LFD equation
with ladder kernel. On the other hand, they are implicitly
included in the three-body BS equation.

The three-body potential, resulted from the first-order
relativistic corrections due to the first eight graphs of
figure~\ref{3bodyf}, was first found in Ref.~\cite{brueckner}. Its
contribution to the triton binding energy was calculated in
Ref.~\cite{yang}. However, in Ref.~\cite{glockle} it was shown
that this relativistic correction cancels with the corresponding
relativistic correction to the second iteration of the one-boson
exchange. Due to this cancellation, sum of these two corrections
does not contribute in the Schr\"odinger equation. However, in
truly relativistic framework, the full graphs figure~\ref{3bodyf}
(not a first-order relativistic correction) should be taken into
account.

The next question is: Can the difference between three-body BS and
LFD results be explained by the diagrams of figure~\ref{3bodyf}?
In order to answer this question, we performed a perturbative
calculation of the correction to the LFD bound state mass
$M_{3,LF}^2$ due to these three-body forces. That is, we
calculated the set of LFD graphs shown in fig.~\ref{3bodyf} and
then found perturbatively the corresponding correction to the LFD
binding energy.

\begin{figure*}[htbp]
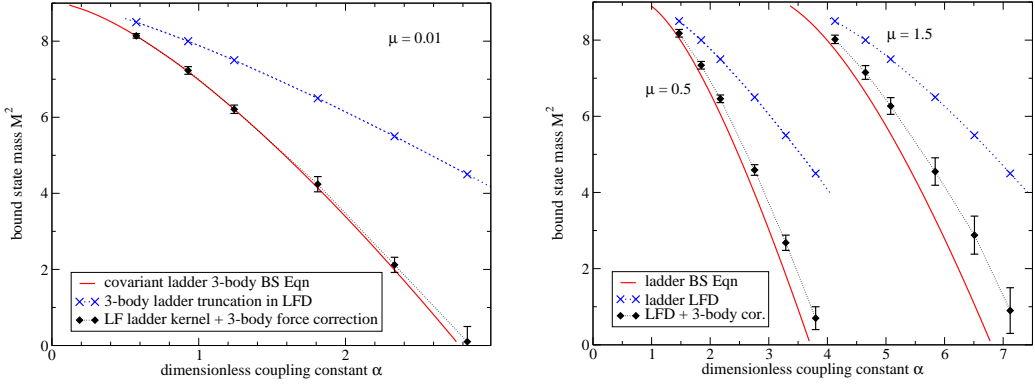

\begin{center}
\begin{minipage}{15.cm}
\mbox{\includegraphics[width=6.5cm]{resfinal_mu0.01.eps}
\hspace{0.5cm}
\includegraphics[width=6.5cm]{resfinal_mu0.5mu1.5.eps}}
\end{minipage}
\end{center}
\caption{Three-body bound state mass squared $M_3^2$ vs. coupling
constant $\alpha=g^2/(16\pi m^2)$ for exchange masses $\mu=0.01$
(top), $0.5$ and $1.5$ (bottom).  The units are set by the
constituent mass: $m=1$.\label{result}}
\end{figure*}

The three-body bound state mass squared $M_3^2$ for the exchange
masses $\mu=0.01$, $0.5$ and $1.5$ are shown in
figure~\ref{result}. Like in figure~\ref{fig5}, the solid and
dotted curves are our results from the three-body BS and LF bound
state equations with the two-body one-boson exchange kernel. As in
figure~\ref{fig5} for binding energy, the BS and LFD masses
squared differ significantly from each other and the difference
increases with increase of the binding energy (and, hence, with
the coupling constant).

The diamonds with error bars in figure~\ref{result} indicate the
sum $M_{3,LF}^2+\Delta M_3^2$, where $\Delta M_3^2$ is the
correction due to the diagrams of figure~\ref{3bodyf}, calculated
perturbatively. The errors indicate the numerical uncertainty,
mainly due to the numerical evaluation of a 12-dimensional
integral $\Delta M^2_3\sim \int \psi K_{123} \psi\ldots$. This
correction does indeed shift the LFD three-body mass squared so
that it comes in reasonably good agreement with the BS results.
This is the effect of relativistic three-body forces, dominated
here by two exchange-bosons in flight, with a small contribution
from the creation (and annihilation) of a pair of constituent
particle-antiparticles.

One should of course keep in mind that here we only present a
perturbative estimate of the correction due to these effective
3-body forces in LFD. Note that the perturbative calculation is
unexpectedly good even for large correction compared to the value
of $M^2$ itself, especially for small values $\mu$ of the mass of
the exchange particle. The agreement becomes less pronounced as
the exchange mass increases, once the exchange mass becomes larger
than the constituent mass. E.g. if the exchange mass is three
times the constituent mass, the perturbative correction explains
about 70\% of the difference in $M_3^2$ between the BS approach
and LFD.

Since the corrections are substantial, one could raise the
question: How reliable is a perturbative calculation in this
context? To answer this question, one can solve three-body LFD
equation beyond the perturbative framework, which requires
significantly more work. Nevertheless, we have ample evidence that
the contributions of effective 3-body forces in LFD are
substantial in the three-body systems, and can explain most of the
difference between the results found in the BS approach and in
LFD.

\section{Conclusion
\label{concl}}
We have solved, for the first time, the three-body BS and LFD
bound state equations with a one-boson exchange kernel and found a
significant difference in the corresponding binding energies. This
difference is absent in the two-body BS and LFD equations
\cite{mc_2000}. After incorporating, in the LFD framework, the
three-body forces of relativistic origin, generated by two
exchange particles in flight, the BS and LFD three-body binding
energies become very close to each other.

In the case of three-nucleon system, there are many other sources
of three-body forces. Part of them is generated by the transitions
$N\leftrightarrow\Delta$. In addition, there can be intrinsic
three-body forces, such as the six-nucleon contact term $NNN\to
NNN$, generated by chiral perturbation theory for the effective
nucleon interaction, or, in QCD, three-body forces due to the
triple gluon vertex. In any case, the two exchange particles in
flight are present and give important contributions. The influence
of such three-body forces on observables significantly exceeds the
experimental precision.

\acknowledgments The authors are sincerely grateful to J.~Vary for
his interest to this work, fruitful discussions and support.  One
of the authors (V.A.K.) is indebted for the warm hospitality of
the nuclear physics group of the Iowa State University (Ames,
USA), where part of the present work was performed. This work was
supported in part by the U.S. Department of Energy Grant
DE-FG02-87ER40371.

\end{document}